\documentclass[sigconf]{acmart}
\usepackage{amsfonts}
\usepackage{amsmath}
\usepackage{natbib}
\usepackage{graphicx}
\usepackage[ruled,linesnumbered]{algorithm2e}
\usepackage{subfigure}
\usepackage{multirow}
\usepackage{color}
\usepackage{hyperref} 
\usepackage{fancyhdr} 
\usepackage{balance}

\AtBeginDocument{%
  }

\copyrightyear{2022}
\acmYear{2022}
\setcopyright{acmcopyright}\acmConference[CIKM '22]{Proceedings of the 31st ACM International Conference on Information and Knowledge Management}{October 17--21, 2022}{Atlanta, GA, USA}
\acmBooktitle{Proceedings of the 31st ACM International Conference on Information and Knowledge Management (CIKM '22), October 17--21, 2022, Atlanta, GA, USA}
\acmPrice{15.00}
\acmDOI{10.1145/3511808.3557403}
\acmISBN{978-1-4503-9236-5/22/10}

\begin{document}
\begin{sloppypar}
\fancyhead[LO]{Multi-Aggregator Time-Warping Heterogeneous Graph Neural Network \\for Personalized Micro-Video Recommendation}
\title{Multi-Aggregator Time-Warping Heterogeneous Graph Neural Network for Personalized Micro-Video Recommendation}

\author{Jinkun Han}
\affiliation{
  \department{Department of Computer Science}
  \institution{Georgia State University}
  \city{Atlanta}
  \state{Georgia}
  \country{USA}
}
\email{ hjinkun1@student.gsu.edu}

\author{Wei Li}
\affiliation{
  \department{Department of Computer Science}
  \institution{Georgia State University}
  \city{Atlanta}
  \state{Georgia}
  \country{USA}
}
\email{wli28@gsu.edu}

\author{Zhipeng Cai}
\affiliation{
  \department{Department of Computer Science}
  \institution{Georgia State University}
  \city{Atlanta}
  \state{Georgia}
  \country{USA}
}
\email{zcai@gsu.edu}

\author{Yingshu Li}
\affiliation{
  \department{Department of Computer Science}
  \institution{Georgia State University}
  \city{Atlanta}
  \state{Georgia}
  \country{USA}
}
\email{yili@gsu.edu}

\renewcommand{\shortauthors}{Jinkun Han, Wei Li, Zhipeng Cai, \& Yingshu Li}
\begin{abstract} 
Micro-video recommendation is attracting global attention and becoming a popular daily service for people of all ages. 
Recently, Graph Neural Networks-based micro-video recommendation has displayed performance improvement for many kinds of recommendation tasks. 
However, the existing works fail to fully consider the characteristics of micro-videos, such as the high timeliness of news nature micro-video recommendation and sequential interactions of frequently changed interests.
In this paper, a novel Multi-aggregator Time-warping Heterogeneous Graph Neural Network (MTHGNN) is proposed for personalized news nature micro-video recommendation based on sequential sessions, where characteristics of micro-videos are comprehensively studied, users' preference is mined via multi-aggregator, the temporal and dynamic changes of users' preference are captured, and timeliness is considered.
Through the comparison with the state-of-the-arts, the experimental results validate the superiority of our MTHGNN model.
\end{abstract}



\maketitle

\section{Introduction} \label{introduction}

Recommendation systems are playing an important role in everyone's daily life~\cite{surveyRS,adomavicius2005toward}. 
In recent years, with the upgraded 4G \& 5G networks, people can watch micro-videos anytime and anywhere, significantly promoting the popularity of micro-videos. 
Two top micro-video companies, TikTok~\cite{tiktokwebsite} and Kwai~\cite{Kwaiwebsite}, recommend micro-videos based on users' interests and profiles~\cite{zhao2015improving}. 
Especially, TikTok has attracted more than one billion monthly-active users around the world~\cite{tiktokuseramount}. 
It is worth mentioning that micro-video recommendation greatly differs from the typical long-video recommendation in terms of logic and needs.

Although micro-videos are attracting more and more global attention, there are only a few works on accurate personalized micro-video recommendation~\cite{ma2018lga, ma2019mmm, 8784862, wei2019mmgcn, jiang2020aspect, cai2021heterogeneous, liu2021concept, lu2021multi, AAAImultiinteraction}. Moreover, the characteristics of micro-videos have not been fully explored yet. 
(i) Compared with long videos, micro-videos can better boost people's social behaviors and encourages people to comment, click, or share, which is ignored by some existing works~\cite{ma2018lga,8784862,cai2021heterogeneous}.
(ii) The interactions in long-video recommendations are usually ``like" or mandatory ratings. Differently, there exist multiple kinds of interactions between users and micro-videos~\cite{AAAImultiinteraction}, including ``like", ``comment", ``forward", ``finish" and so on.
(iii) A micro-video is typically composed of multi-modal, such as music, title, tags, video content, and special effects, which is one of the specialties of micro-videos. 
%
(iv) In micro-video recommendation, capturing frequently changed preference is important because people's interests are influenced by historical interests, current viewing contents, and hot topics, and thus frequently changed over time, which is not studied by many current works~\cite{ma2018lga,ma2019mmm,wei2019mmgcn,cai2021heterogeneous}.
%
(v) Micro-videos require high timeliness as burgeoning news delivery and entertainment service as people usually prefer new micro-videos; that is, people would like to view the latest news, social hot spots, or the latest trends of the producers they followed, which has not been addressed by the existing works~\cite{ma2018lga, ma2019mmm, 8784862, wei2019mmgcn, jiang2020aspect, cai2021heterogeneous, liu2021concept, lu2021multi, AAAImultiinteraction}.

Motivated by the above observations, in this paper, we aim to develop a new micro-video recommendation system framework that can comprehensively exploit the aforementioned characteristics of micro-videos with enhanced recommendation accuracy.
To this end, our Multi-aggregator Time-warping Heterogeneous Graph Neural Network (MTHGNN) is proposed for personalized news nature micro-video recommendation based on sequential sessions. 
Technically speaking, the innovation of our MTHGNN model lies in the following four aspects.  
(i) To fully depict the characteristics of strong social behaviors and multi-modals, a directed time-warping heterogeneous graph is constructed to build the connections of users, micro-videos, and multi-modals. The edges on the heterogeneous graph are marked with different relation types, including ``finish", ``like", ``non-interaction", {\em etc.}, to simulate the multi-type interactions in micro-videos. 
(ii) To better extract embeddings of users' preference, multi-aggregator algorithm~\cite{multi-aggregator} is applied to embed the complex heterogeneous information on nodes, in which Relational Heterogeneous Message Passing aggregator and Attention Heterogeneous Message Passing aggregator are deployed.
(iii) To characterize the temporal and dynamic changes of the users' preference for micro-videos, we divide the multi-modal heterogeneous graph into a series of sequential sessions based on the timestamps of interactions to fuse historical interests into the latest interests. 
Results from the prior work~\cite{groupgraph} suggest that groups that gather the same preference lead to better performance. Hence, we consider sessions as groups with similar features of the interacted micro-videos.  
(iv) To recommend fresh micro-videos, timeliness is considered from  multi-modal heterogeneous graph, LSTM aggregator, and sequential sessions.

In a summary, the multi-fold contributions of this paper are addressed below.
\begin{itemize}
\item To the best of our knowledge, this is the first work that investigates multiple perspectives of the characteristics of micro-videos while inventively enhancing micro-video timeliness for personalized recommendation.

\item A novel model termed MTHGNN is proposed to recommend people personalized news nature micro-videos, taking into account the characteristics of micro-videos for recommendation performance improvement. 

\item Extensive experiments are well set up on two real-world datasets of TikTok and MovieLen for performance evaluation, and the results demonstrate that our MTHGNN outperforms the state-of-the-art models not only on micro-video recommendation, but also on long-movie recommendation.
\end{itemize}

The rest of this paper is organized as follows.
We brief the related works on micro-video recommendation in Section~\ref{RelatedWork}.
After presenting the problem statement in Section~\ref{ProblemStatement}, we propose our recommendation model in Section~\ref{Methodology}.
In Section~\ref{Results}, we evaluate our proposed methods through real-data experiments.
Finally, this paper is concluded in Section~\ref{Conclusion}.

\section{Related Works} \label{RelatedWork}

Micro-video has become more and more popular in recent years. 
In 2018, Ma {\em et al.}~\cite{ma2018lga} proposed a micro-video recommendation model based on a deep neural network exploiting visual and textual information.
Ma {\em et al.}~\cite{ma2019mmm} and Liu {\em et al.}~\cite{8784862} started to combine collaborative filtering with multi-modals to recommend micro-videos. 
While, many of the existing works since 2019 focus on graph-based methods.
Wei {\em et al.}~\cite{wei2019mmgcn} proposed a graph neural network model, MMGCN, which constructs a heterogeneous graph to extract homogeneous information of multi-modals. 
In~\cite{jiang2020aspect}, time effects on different users were studied to develop MTIN that uses session-based video list and mines user group interests to recommend micro-videos. 
Furthermore, in 2021, researchers started to explore GNN information from the aspects of the contributions of multi-modals, purifying graphs, interest groups, and sequence of interactions. 
Cai {\em et al.}~\cite{cai2021heterogeneous} proposed a heterogeneous GNN model, termed HHFAN, which utilizes a hierarchical structure to compute the contributions of multi-modals. 
CONDE was introduced as a way to integrate textual features and GNN to facilitate video content mining and user preference generation, as well as a denoising procedure to eliminate noisy concepts and poor clicks~\cite{liu2021concept}.  
Yao {\em et al.}~\cite{AAAImultiinteraction} tried to consider the influence of multi-type interaction.
Lu {\em et al.}~\cite{lu2021multi} proposed DMR to obtain various and dynamic preferences based on historical interests and trend groups. 
Some of the above work considered the temporal user-to-micro-video interactions, while some took into account group interests. 
But, none of them address the social relations and behaviors between users, users to micro-videos interactions, micro-videos to multi-modal interactions, or the multi-type user-micro-video interactions. 

To enhance the performance of micro-video recommendation, in this paper, we propose a novel model, MTHGNN, to comprehensively exploit the characteristics of complex social relations by computing relation weights, multi-type interactions between users and micro-videos, session-based short-time interests, sequential session-based long-time feature extraction, and high timeliness of micro-videos.

\begin{figure*}
    \centering
    \includegraphics[width=0.95\textwidth]{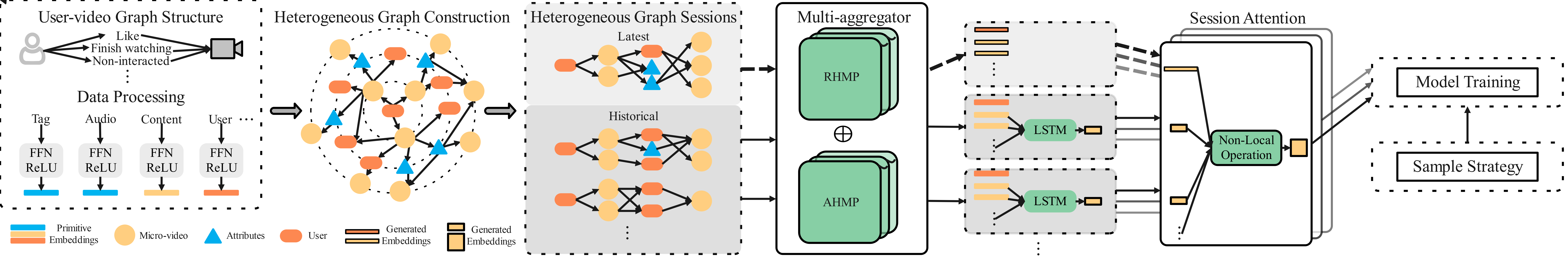}
    \caption{The Framework of our proposed MTHGNN ($\oplus$ represents concatenation operation).}
    \label{fig:system_framework}
\end{figure*}

\section{Problem Statement} \label{ProblemStatement}
Formally, the set of users is denoted as $U$, and the set of micro-videos is denoted as $V$.
Each micro-video owns its multi-modal attribute features, such as the titles of the micro-videos, the tags of the events, and the background music used by the micro-video producers, which are represented as $V_{attrs}=\left\{v_1^A,v_2^A,v_3^A,\ldots,v_{\left|V\right|}^A\right\}$ with $A=\left \{ Tag, Title, Audio, \ldots \right \}$. 
Generally speaking, a user may sequentially interact with micro-videos during different time periods. 
The matrix of user-video interactions is denoted by $O^{\left| U \right| \times \left| V \right|}$ based on the implicit interactions provided by the users, where for $u\in U$ and $v \in V$, $o_{uv} \in \{0, 1\}$ is equal to 1 only if user $u$ interacts with micro-video $v$.
In this paper, given any user $u$ with a set of historical interacted micro-video $V^{u}$, our algorithm MTHGNN is expected to output the user preference and micro-video embeddings, including the latest and historical information of user interests. 
First of all, for each user $u$, MTHGNN selects $M^{u}$ consecutive micro-videos to construct $u$'s heterogeneous graph $G^{u}$, in which the micro-videos interacted with $u$ are sorted based on their timestamps. 
Next, MTHGNN sequentially selects $m$ interacted micro-videos to form a session, which can represent a user's short-time preference. 
A series of sessions is defined as sequential sessions. 
Accordingly, user $u$'s sequential session can be expressed as $S^u=\left \{ s_1^u, s_2^u, \ldots, s_{\pi^{u}}^u \right \}$, where $\pi^{u}\leq \pi_{max}$ and $\pi_{max}$ is the pre-determined maximum length of all session sets.
Particularly, the set $S_{\phi}=\left \{ s_{\pi^{u}}^u \right \}$ represents the latest session used to mine the latest user interests, and the set $S_{\psi}=\left \{ s_1^u, s_2^u, \ldots, s_{{\pi^{u}}-1}^u \right \}$ represent the historical sequential sessions used to extract long-term user preferences. 
Finally, when MTHGNN terminates, the predicted scores $\left \{ \widehat{o} _{uv_{1}},\widehat{o} _{uv_{2}}, \ldots , \widehat{o} _{uv_{i}} \right \}$ for user $u$ are computed for top-$K$ recommendation by using the extracted embeddings, which is detailed in Section~\ref{Methodology}.

\begin{figure}
\centering
    \subfigure[Homogeneous graph]
    {\includegraphics[width=0.18\textwidth]{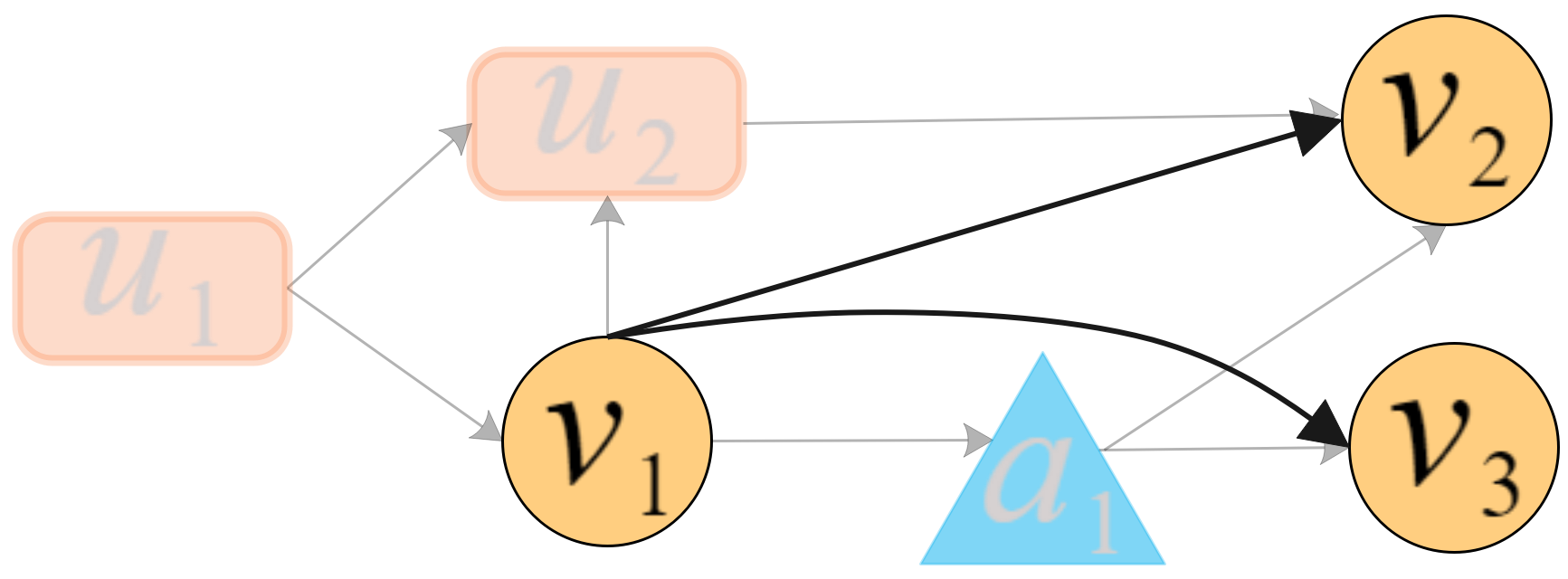}
    \label{Homogeneousgraph}} 
    \hfil
    \subfigure[Multi-modal heterogeneous graph]
    {\includegraphics[width=0.18\textwidth]{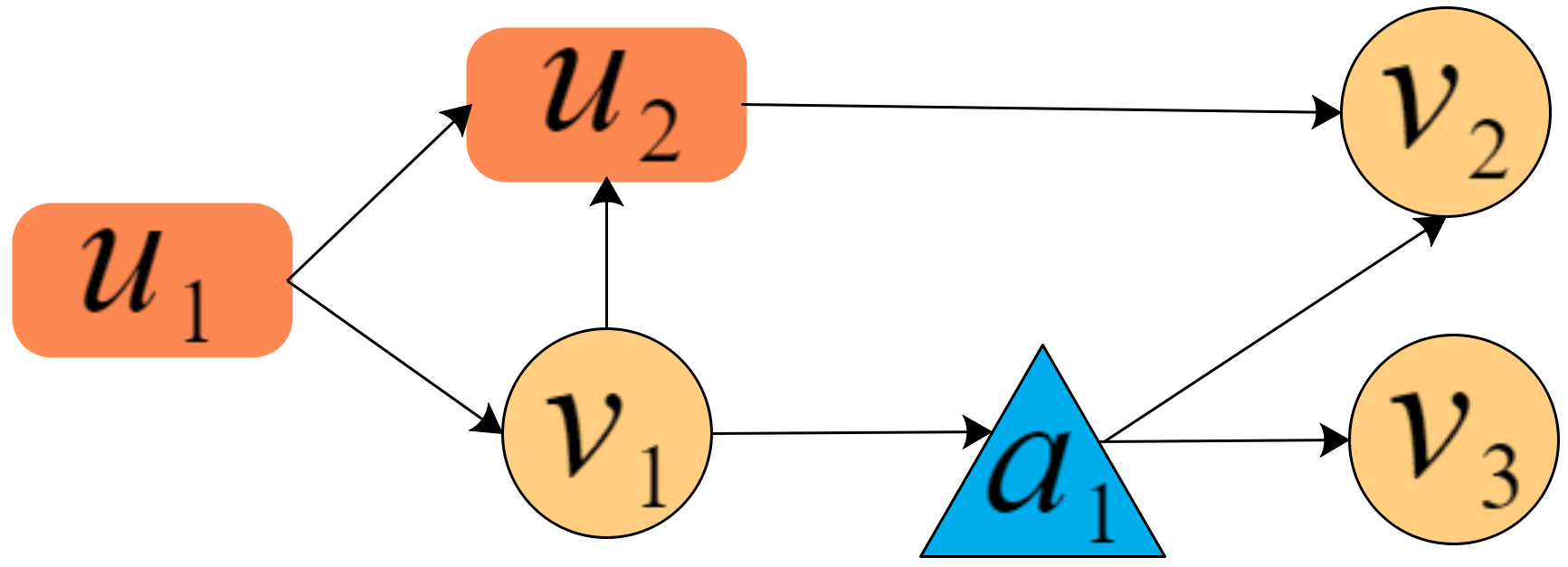}
    \label{Heterogeneousgraph}} 
\caption{Homogeneous graph vs. multi-modal heterogeneous graph}
\label{fig:Homogeneous_Heterogeneous_graph}
\end{figure}

\section{Methodology} \label{Methodology}

In this paper, a novel  Multi-aggregator Time-warping Heterogeneous Graph Neural Network (MTHGNN) is proposed to solve the challenges described in Section~\ref{introduction}. As shown in Fig.~\ref{fig:system_framework}, our proposed model framework consists of seven components, including: (i) User-video Graph Structure and Data Processing, (ii) Heterogeneous Graph Construction, (iii) Heterogeneous Graph Sessions, (iv) Multi-aggregator, (v) Session Attention, (vi) Sample Strategy, and (vii) Model Training \& Prediction. The design of these components in MTHGNN are demonstrated as follows.

\begin{algorithm}
\SetAlgoLined
\caption{
Multi-modal Time-warping Heterogeneous Graph Construction}

\label{alg:BFS}

 \KwIn{the target user $u$, 
 the set of historical interacted micro-video $V^{u}$,
 the number of selected historical interacted micro-video $M^{u}$, 
 the  number of interacted micro-videos in each session $m$, 
 the number of selected neighbors $m'$, 
 the node set $N^l$ contains all the nodes in layer $l$,
 the number of layers $h$}
 \KwOut{the set of directed subgraphs $g$, which contains latest and historical sessions}
   Initialization: $N^{0}=\left \{ u \right \}$, add node $u$ into graph $G^u$, select $M^u$ historical interacted micro-videos from $V^{u}$, which are  ordered by timestamp, and update $N^{1}=\left \{ v_{1}...v_{M^{u}} \right \}$\;
  \For{$v$ in $N^{1}$}
     {
        Build edge $u \rightarrow v$ in $G^u$, and mark the edge type \;
     }
  \For {$l$=2...$h$}  
     {
         Temporary set of nodes $N_{tmp_{1}}=\left \{  \right \}$\;
        {
            \For{$n$ in $N^{l-1}$}
            {
                Temporary set of nodes  $N_{tmp_{2}}=\left \{  \right \}$\;
                Select node $n$'s atrributes as nodes, and add them into $N_{tmp_{2}}$\;
                Select $m'$ latest interacted users \textbf{if} $n$ is a micro-video node and more than $m'$ interacted users are found, and add into $N_{tmp_{2}}$\;
                Select $m'$ latest interacted micro-videos \textbf{if} $n$ is a user node and more than $m'$ interacted micro-videos are found, and add into $N_{tmp_{2}}$\; 
                \For{$n_{neighbor}$ in $N_{tmp_{2}}$}
                 {
                     Build edge $n \rightarrow n_{neighbor}$ in $G^u$, and mark the edge type\;
                 }
                $N_{tmp_{1}} = N_{tmp_{1}} \cup N_{tmp_{2}}$\;
            }
            
        }
         $N^{l} = N_{tmp_{1}}$\;
     }
   Session index $i = 0$\;
  \While{$(i+1) \times m \leq  M^{u}$} 
       {
            $g_{s_{i}} = branches(G^u, i\times m, (i+1) \times m)$ \;  
            $g = g \cup g_{s_{i}}$, $i = i+1$\; 
       }
  \KwResult { $g$}
\end{algorithm}

\subsection{User-video Graph Structure and Data Processing}
The interaction type between users and micro-videos is helpful to mine user's preference patterns. 
For example, if user $u_1$ likes micro-video $v_1$ and finishes watching micro-video $v_2$, $v_1$ contributes more information to the user's preference, because ``like" is an explicit preference that the user tells us directly, while ``finish watching" is an implicit preference implying the user may be interested in the video. However, the traditional models do not utilize such latent information. 

To address this problem, the interaction type is taken into account in the relation defining layer, {\em i.e.}, the multi-relation is defined as 
$R=\left \{r_{\iota}^{v2u}, r_{\iota}^{u2v}, r_f^{v2u}, r_f^{u2v}, r_{p}^{u2v}, r_{p}^{v2u}, r^{a2v}, r^{v2a} \right \}$, where $r_{\iota}^{v2u}$ and $r_{\iota}^{u2v}$ represent the relations that user $u$ likes micro-video $v$, $r_f^{v2u}$ and $r_f^{u2v}$ denote the relations that user $u$ finishes watching micro-video $v$, $r_{p}^{u2v}$ and $r_{p}^{v2u}$ mean that micro-video $v$ is produced by user $u$, and $r^{a2v}$ and $r^{v2a}$ ($a\in A$) reveal how micro-video $v$ is connected with its attributes.
In this way, the kinds of paths and relations between nodes can be employed to construct the multi-modal heterogeneous graph.

The data processing layer initializes the embedding of users, micro-videos, and attributes via one-hidden-layer feedforward network (FFN) with the same embedding length. The design of FFN allows the model to produce new embeddings when facing unseen or new-coming nodes, so that the model can effectively work even when some new users and/or micro-videos join in this system. Rectified Linear Unit (ReLU)~\cite{nair2010rectified, maas2013rectifier} is mixed with the FFN network to increase the nonlinear capability of the proposed model.
%


\subsection{Heterogeneous Graph Construction and Heterogeneous Graph Sessions} \label{HGCHGS}
For effectively modeling the information of the users, micro-videos, and multi-modal attributes, Multi-modal Heterogeneous Graph (MHG) is used to represent the connections and relations between nodes. 
The global MHG is formally denoted as $G=(N,E,R)$, where $N=U\cup V\cup V_{attrs}$ is the set of multi-type nodes, and 
$E=\left \{ e_{i,j}|i \in N, j \in N  \right \}$ is the set of edges. If there is a kind of relation between nodes $i$ and $j$, an edge $e_{i,j}$ is connected between node $i$ and $j$ on graph $G$ with their corresponding relation type $r \in R$.
To improve the time efficiency of training process, for each user $u$, the local MHG $G^{u}$ is constructed in the Heterogeneous Graph Construction component as presented in Algorithm~\ref{alg:BFS}, and then the global graph is only used in the sampling components as described in Section~\ref{Sampling_Strategy}.

The purpose of constructing $G^{u}$ is to mine the target user's latent preference carved by the micro-videos, attributes, producers, and other users.
As shown in Fig.~\ref{fig:Homogeneous_Heterogeneous_graph}, compared with a homogeneous graph, MHG reflects more social relations with a hierarchical graph structure. From Fig.~\ref{Homogeneousgraph}, the importance of micro-videos $v_{1}$ and $v_{2}$ cannot be clearly distinguished since the nodes on the homogeneous graph does not consider the interactions of the other types of nodes. While, through our MHG in Fig.~\ref{Heterogeneousgraph}, it can be seen that the importance of micro-video $v_{2}$ is higher than that of $v_{1}$ because three paths directs $u_{1}$ to $v_{2}$. 
Therefore, compared with the existing graph models mentioned in related works, the MHGs of our MTHGNN can keep the original structural and modal information in the constructed graphs.

Given a target pair $(u,V^{u})$, $M^{u}$ interacted micro-videos are selected to construct the initial MHG $G^{u}$. 
By dividing the initial MHG $G^{u}$ into the latest and historical sessions, each session is extracted with $m$ interacted micro-videos. 
Finally, $\pi^{u}$ sessions are reconstructed, each of which can be regarded as a community that gathers users who have the same implicit tastes depicted by similar background music, tags, contents, or producers.
To explore the high-order relationships among the users, the micro-videos, and the attributes, the number of layers is utilized to determine the size of the constructed MHG. 
The layer that the center node locates is defined as layer 0, and the other layers are numbered as $l\in \left \{1, 2,..., h\right \}$ based on the number of hops, $h$, from the center node. 
In the initial MHG $G^{u}$, the center node is the target user $u$; while in the component of Heterogeneous Graph Sessions, the center node can be the target $u$ or any interacted micro-video $v$ of $u$'s sessions.
As presented in Algorithm~\ref{alg:BFS}, for the target user $u$, $G^u$ is built based on the breath-first search (BFS) algorithm in lines 1-18 and is divided into sequential sessions in lines 19-23.
Function $branches(graph, start\_index, end\_index)$ returns a subgraph that contains the branches between $start\_index$ and $end\_index$.
Finally, a set of subgraphs, $g=\left \{ g_{s_{1}},g_{s_{2}},\ldots,g_{s_{\pi^{u}}} \right \}$, are obtained in the Heterogeneous Graph Sessions.

\begin{algorithm}
\caption{Generation of Session Embeddings of Target User and Micro-videos}
\label{alg:reverse_user_video_edge}

\KwIn{subgraph $g_{s_{i}}$, the target user $u$, the set of interacted micro-video  $\varphi_{g_{s_{i}}}(u)$}
\KwOut{ the set of user $u$ and $u$'s interacted micro-videos embeddings  $z_{g_{s_{i}}}$ }
  Execute Eq.~\eqref{eq:relation}, Eq.~\eqref{eq:attention_1}, Eq.~\eqref{eq:relation_attention_concatenated} to generate $z_{g_{s_{i}}}^{u}$
  $z_{g_{s_{i}}} = z_{g_{s_{i}}} \cup z_{g_{s_{i}}}^{u}$\;
  \For{$v$ in $\varphi_{g_{s_{i}}}(u)$}
     {
         Reverse directed edge $u\rightarrow v$ of $g_{s_{i}}$ to $v\rightarrow u$\;
         Add the edge type of $v$ to $u$ and set $v$ as target node\;
         Execute Eq.~\eqref{eq:relation}, Eq.~\eqref{eq:attention_1}, Eq.~\eqref{eq:relation_attention_concatenated}\;
         $z_{g_{s_{i}}} = z_{g_{s_{i}}} \cup z_{g_{s_{i}}}^{v}$\;
     }
\KwResult{$z_{g_{s_{i}}}$ where $z_{g_{s_{i}}} = \left \{z_{g_{s_{i}}}^{u}, z_{g_{s_{i}}}^{v_{1}},\ldots,z_{g_{s_{i}}}^{v_{m}} \right \}$}
\end{algorithm}

\subsection{Multi-aggregator}
To extract the complex relationships and information from the interactions between the users and the micro-videos, a Relational Heterogeneous Message Passing (RHMP) aggregator and an Attention Heterogeneous Message Passing (AHMP) aggregator are developed to generate the user's preference embeddings and micro-video representations inspired by the work on multi-aggregator~\cite{multi-aggregator} and advanced aggregators~\cite{busbridge2019relational, velivckovic2017graph, hsu2021retagnn}. 
With our constructed MHG as the input, and the outputs of RHMP and AHMP are the embeddings of the target user and the interacted micro-videos.

RHMP starts from the target user $u$ or anyone of $u$'s interacted micro-video $v$ of MHG $g_{s_{i}}$, where $g_{s_{i}} \in g$. 
To clearly show the connections between nodes of two adjacent layers, $N_{g_{s_{i}}}^{l}$ denotes the node set at layer $l$ of subgraph $g_{s_{i}}$. 
While $\varphi_{g_{s_{i}}}^{l} \left ( n \right ) \subseteq  N_{g_{s_{i}}}^{l}$ is the set of the one-hop neighbors of node $n$ at layer $l$ of $g_{s_{i}}$ with $n \in N_{g_{s_{i}}}^{l-1}$.
The center embedding of layer $l$ is denoted as $x_{g_{s_{i}}}^{*,l} \in \mathbb{R}^{1\times d}$, where $*$ is either $u$ representing the target user $u$ or $v$ representing $u$'s interacted micro-video $v$. 
To update the center embedding of relational information of layer $\left (l+1\right )$, a two-step RHMP is applied:
\begin{equation}
\label{eq:relation}
x_{g_{s_{i}}}^{*,l+1}=x_{g_{s_{i}}}^{*,l}W_{0}^{l}+\sum_{n\in N_{g_{s_{i}}}^{l-1}}\sum_{j\in\varphi_{g_{s_{i}}}^{l}(n)}\sum_{r\in R} x_{j}^{l}W_{r}^{l},
\end{equation}
where $W_{0}^{l}\in \mathbb{R}^{d \times d}$ and $W_{r}^{l} \in \mathbb{R}^{d \times d}$ are the transform matrices of learning parameters, $x_{j}^{l}$ represents the embedding of node $j$, $l\in \left \{1, 2,..., h\right \}$.
The RHMP consists of two parts, including relational tracing and message passing mechanism. 
Firstly, in the relational tracing part, the relation of each node is traced by $x_{j}^{l}W_{r}^{l}$, where the information of each node is transformed into a relational space based on $r$. 
The users have multi-type interactions with micro-videos, such as like, finish watching, comment, and so on.
Although all of them are regarded as interactions, they contribute differently to the user's preference. 
The relational tracing matrices help the model to distinguish the contributions from different relations. 
Secondly, in the message passing part, the summation of all transformed neighbor nodes $x_{j}^{l}$ is combined with 
the center embedding $x_{g_{s_{i}}}^{*,l}$, where the relation matrix $W_{0}^{l}$ represents how much center information is kept from layer $l$ to layer $l+1$.

The aggregator AHMP calculates attention weights to reflect the importance of nodes towards to the corresponding center embedding, which can be expressed by Eq.~\eqref{eq:attention_1} and Eq.~\eqref{eq:attention_2}.
\begin{equation}
\label{eq:attention_1}
y_{g_{s_{i}}}^{*,l+1}=y_{g_{s_{i}}}^{*,l}W_{0}^{l}+\sum_{n\in N_{g_{s_{i}}}^{l-1}}\sum_{j\in \varphi_{g_{s_{i}}}^{l} (n)} \alpha _{j}^{l}y_{j}^{l},
\end{equation}
\begin{equation}
\label{eq:attention_2}
\alpha _{j}^{l} = softmax\left ( P_{\theta }\left ( Q_{\theta }^{l}\left ( y_{g_{s_{i}}}^{*,l}\oplus y_{j}^{l} \right )^{T} \right ) \right ),
\end{equation}
where $y_{g_{s_{i}}}^{*,l+1}$ the center embedding at layer $l+1$, $\alpha _{j}^{l}$ is the attention weight, $Q_{\theta }^{l}\in \mathbb{R}^{1 \times 2d}$ denotes a learnable matrix, $P_{\theta }$ is the LeakyRelu function.

After computing the center embedding of layer $h$, RHMP and AHMP generate the final embedding $x_{g_{s_{i}}}^{*,h}$ and $y_{g_{s_{i}}}^{*,h}$, which are concatenated to represent a user preference or a micro-video representation in a session:
\begin{equation}
\label{eq:relation_attention_concatenated}
z_{g_{s_{i}}}^{*} = x_{g_{s_{i}}}^{*,h} \oplus y_{g_{s_{i}}}^{*,h}.
\end{equation}
Hence, each session generates the target user $u$'s embedding $z_{g_{s_{i}}}^{u}$ and $m$ interacted micro-video embeddings $\left \{ z_{g_{s_{i}}}^{v_{1}},z_{g_{s_{i}}}^{v_{2}}, \ldots, z_{g_{s_{i}}}^{v_{m}}\right \}$ via Algorithm~\ref{alg:reverse_user_video_edge}. 
First of all, Algorithm~\ref{alg:reverse_user_video_edge} executes line 1 to generate $z_{g_{s_{i}}}^{u}$. 
Then, lines 2-7 are applied to reverse the center node from the target user $u$ to the interacted micro-videos and generate $z_{g_{s_{i}}}^{v}$. Finally, the set of embeddings of the session $s_i$ is returned.

There are $\pi^{u}$ sessions that are divided as the latest session in $S_{\phi}$ and the historical sessions in $S_{\psi}$ so that the historical micro-video embeddings of each session can be merged for the session attention component. 
In this paper, the Long Short-Term Memory (LSTM) aggregator~\cite{hamilton2017inductive} is applied to aggregate the interacted sequential micro-videos in each session:
\begin{equation}
\label{eq:LSTM}
z_{g_{s_{i}}}^{\sigma } = \frac{1}{\left| \varphi_{g_{s_{i}}} \left ( u \right ) \right |}\sum_{v\in \varphi_{g_{s_{i}}} \left ( u \right )}^{}LSTM\left ( z_{g_{s_{i}}}^{v} \right ),
\end{equation}
where $\varphi_{g_{s_{i}}} \left ( u \right )$ is the set of all one-hop neighbors of the target user $u$ in the subgraph $g_{s_i}$, and $s_{i} \in S_{\psi}$.
LSTM is used to mine the interest trend of each session, implying which micro-videos the target user is going to like after each session. Such trend mining also ensures the timeliness of recommendations.
%

\subsection{Session Attention} \label{Session_Attention}
Micro-video recommendation is highly time-dependent. With the news happening in the society, the users may be interest in the most recent micro-videos corresponding to the user's preference. 
Notice that these $M^{u}$ interacted micro-videos divided into $\pi^{u}$ sessions can reflect users' preference to a certain extent. 
In the session attention component, the non-local operation~\cite{wang2018non} is used to seek user's historical preference from the historical local sessions, which can be performed by Eq.~\eqref{eq:session_attention_1} and Eq.~\eqref{eq:session_attention_2}.
\begin{equation}
\label{eq:session_attention_1}
z_{u}^{{v}}=\frac{1}{\sum_{s_{i}\in  S_{\psi}}  [z^{v}_{g_{s_{\pi^u}}}\odot (z^{\sigma}_{g_{s_{i}}})^{T}] } \sum_{s_{i}\in  S_{\psi} } [z^{v}_{g_{s_{\pi^u}}}\odot (z^{\sigma}_{g_{s_{i}}})^{T}] \times [z^{\sigma}_{g_{s_{i}}}W'],
\end{equation}
\begin{equation}
\label{eq:session_attention_2}
z_{u}^{u}=\frac{1}{\sum_{s_{i}\in  S_{\psi} }  [z^{u}_{g_{s_{\pi^u}}}\odot (z^{u}_{g_{s_{i}}})^{T}] }\sum_{s_{i}\in  S_{\psi} }  [z^{u}_{g_{s_{\pi^u}}}\odot (z^{u}_{g_{s_{i}}})^{T}] \times [z^{u}_{g_{s_{i}}}W'],
\end{equation}
 where $(\cdot)^{T}$ denotes the transpose of the input, and $W' \in \mathbb{R}^{2d \times 2d}$ is a learnable matrix.
 In the original design of non-local network~\cite{wang2018non}, the similarity is calculated by $exp(\cdot \odot  \cdot)$.
Since we use $Sigmoid(\cdot)$ function in the optimization component of our MTHGNN mechanism, the combination of $exp(\cdot \odot \cdot)$ and $Sigmoid(\cdot)$ may cause the gradient disaster. 
Thus, in this paper, $exp(\cdot \odot  \cdot)$ is replaced by $(\cdot \odot  \cdot)$ to avoid scaling the similarity. 
In the session attention component, $m$ embeddings of the latest interacted micro-video are used to fuse the historical information of each session based on the similarities generated by $(\cdot \odot  \cdot)$. 
Hence, MTHGNN can recommend a latent micro-video only when it has a high similarity with the latest $m$ micro-videos and also meets a user's historical preference. 
Finally, Session Attention component generates the set of embeddings $\left \{ z_{u}^{u},z_{u}^{v_1},z_{u}^{v_1},\ldots,z_{u}^{v_m} \right \}$ for Model Training component.

\subsection{Sampling Strategy} \label{Sampling_Strategy}
The traditional graph-based micro-video recommendation~\cite{wei2019mmgcn, cai2021heterogeneous} generates micro-video embeddings for optimization or prediction via graph training, in which we have to retrain the model if there is one new-coming micro-video or unseen micro-video.
Differently, we develop a graph-free method to generate micro-video embeddings for optimization and prediction, which can deal with unseen nodes and new-coming nodes.
Our sampling strategy only considers the vectors of any sampled micro-video $v_{spl}$ and its neighbors; that is,
\begin{equation}
\label{eq:sampling_neighbors}
z^{\omega} = \frac{1}{\left |\varphi_{G}(v_{spl}, \omega)  \right | } \sum_{i\in \varphi_{G}(v_{spl}, \omega)} x_{i},
\end{equation}
where $\varphi_{G}(v_{spl}, \omega)$ is the set of $v_{spl}$'s one-hop neighbors that are of the type $\omega$ on the graph $G$, and 
 $\omega \in N^{type} = \left \{User, Video, Producer, Tag, Audio,... \right \}$ represents the node type, and $x_{i}$ is the primitive embedding of node $i$.

When processing $\varphi_{G}(v_{spl}, User)$, the vector of the target user should be removed because the recommendation system is not able to know whether there is an interaction between the target user and the sampled micro-videos in advance. Hence, the user set of neighbors of $v$ is $\varphi_{G}(v_{spl}, User) \setminus u$.
\begin{equation}
\label{eq:sampling_meanwithMC}
z_{u} ^{v_{spl}} =\frac{1} { | N^{type} | } \sum_{\omega \in N^{type} } z^{\omega},
\end{equation}
where $z_{u}^{v_{spl}}$ is the final aggregated embedding of the current sampled micro-video for the target user $u$. 
The graph-free sampling strategy is utilized to aggregate the information from different kind of nodes, which helps the model learn the weights of relations and attention.

On the other hand, the graph-free sampling strategy provides strong adaptive capability for real-world scenarios. In real-world recommendation, generating embeddings via graphs and model wastes time and requires abundant computing resources. 
Moreover, the graph-free sampling strategy only stacks raw vectors, which almost takes $O(1)$ complexity; while the complexity of graph-based sampling depends on the structure of recommendation models.

\subsection{Model Training \& Prediction}
In the model training and prediction component, we use Eq.~\eqref{eq:BPR_score} to compute the predicted score for each sampled micro-video and rank the top-$K$ micro-videos for recommendation.
\begin{equation}
\label{eq:BPR_score}
\begin{aligned}
    \widehat{o} _{uv} = z_{u}^{u}\odot z_{u} ^{v_{spl}}+\sum_{v \in s_{\pi^{u}}^{u}}^{} z_{u}^{v}\odot z_{u} ^{v_{spl}}.
\end{aligned}
\end{equation}
Specifically, $\widehat{o} _{uv}$ indicates the similarity between the target user $u$ and the users who like $v_{spl}$ as well as the similarity between $v_{spl}$ and $u$'s latest historical interests.
If the sampled micro-video satisfies the user's interests, the predicted score $\widehat{o} _{uv}$ should be high; otherwise, a low score is given.

Furthermore, MTHGNN is optimized by making the gap of the scores between liked and disliked micro-video maximized.
To this end, BPR loss function~\cite{BPR} is applied for personalized micro-video recommendation. The overall loss function includes BPR loss function, relation limitation, and parameter regularization, as shown in the following.
\begin{equation}
\label{eq:BPR_loss_function}
\begin{aligned}
    loss&=\frac{1}{\left | U \right |} \sum_{u\in U}\sum_{v \in V_{p} }\sum_{v' \in V_{n} }-log Sigmoid \left ( \widehat{o}_{uv} - \widehat{o}_{uv' } \right )\\
    &+\beta \left ( \sum_{r\in R}\sum_{l=0}^{h-1} \left \| W_{r}^{l+1}-W_{r}^{l}  \right \|_{F}^{2}   \right ) +\eta \left \| \theta  \right \|_{F}^{2},
\end{aligned}
\end{equation}
in which $ V_{p}$ is the set of interacted micro-videos, $V_{n}$ is the set of non-interacted Micro-videos, $\beta$ and $\eta$ are the regularization weights, 
$Sigmoid(\cdot)$ is the Sigmoid function,
$\theta$ contains all the learnable parameters of the model including the parameters in FFN, $W_{0}^{l}$, $W_{r}^{l}$, $Q_{\theta }^{l}$, and $W'$,
and $\left \| \cdot  \right \| _{F}^{2} $ represents the Frobenius norm. 
The relation limitation part of the loss function limits the same relation to be similar. 
If there is a large gap between the transform matrices $W_{r}^{l+1}$ and $W_{r}^{l}$, the transformed node embeddings of different relations at the same layer might be similar, which causes the graph learning to be over-smooth, so the second term is considered in the loss function.

\section{Performance Evaluation} \label{Results}

In this section, a series of experiments are conducted to evaluate the performance of our proposed MTHGNN mechanism.

\subsection{Datasets}
We conduct experiments on two real-world datasets, including MovieLen dataset and TikTok dataset, which are described below with the statistics summarized in Table~\ref{table:dataset}.

\textbf{(i) TikTok~\cite{TikTokdataset}:} TikTok dataset is collected from real-world anonymized users and contains the interactions of ``like", ``finish", and ``Non-interacted"  between users and micro-videos. In our experiments, the interactions of ``like" and ``finish" are marked as interacted, while the interactions of ``non-interaction" is marked as non-interacted.
To exam the performance of recommendation models under various application scenarios with different interaction densities (see Table~\ref{table:dataset}), we randomly sample TikTok(1/50) and TikTok(1/5) for TikTok dataset based on sampling rate of 0.02 and 0.2, respectively.

\textbf{(ii) MovieLen~\cite{MovieLendataset}:} MovieLen contains the ratings of movies rated by real-world users, which is widely used in recommendation systems for performance evaluation. 
MovieLen contains the rating scores from 1 to 5. 
As BPR loss function ({\em i.e.}, Eq.~\eqref{eq:BPR_loss_function}) requires implicit data clearly reveals whether a user likes a micro-video or not, we set the score of 4 as unclear status (the threshold), where the rating of 5 is marked as interacted ({\em i.e.}, a user likes a micro-video), and the ratings of 0 to 3 are marked as non-interacted ({\em i.e.}, a user dislikes a micro-video). 
\begin{table}[!t]
\caption{Statistics of datasets ($Density=\frac{Interactions}{Users\times Videos}$).}
\begin{center}
\scalebox{0.85}{\begin{tabular}{c c c c c c}
    \toprule
    \textbf{Dataset} &\textbf{Type} & \textbf{Users} & \textbf{Videos} & \textbf{Interactions} & \textbf{Density}  \\
    \midrule
      TikTok(1/50) & Micro-video & 1434  & 29662  & 95,426    &  0.224\%  \\ 
      TikTok(1/5)  & Micro-video & 16538 & 366,017& 1,047,358 &  0.017\%  \\
      MovieLen     & Movie       & 6040  & 3884   & 1,000,209 &  4.264\%  \\
    \bottomrule
\end{tabular}}
\end{center}
\label{table:dataset}
\end{table}

\subsection{Baseline Models}
The following state-of-the-art recommendation models are adopted for a comprehensive comparison.

\textbf{(i) GraphRec~\cite{fan2019graph}} is a framework that recommends people goods based on social relations and group interests.

\textbf{(ii) SASRec~\cite{kang2018self}} processes sequential interactions and computes the relevance of item pairs for recommendation.

\textbf{(iii) MMGCN~\cite{wei2019mmgcn}} is a GCN-based multi-modal framework that generates representations of users and micro-videos by capturing textual, acoustic, and visual features. 
    
\textbf{(iv) R-GCN~\cite{rgcn}} can be used for recommendation by constructing complete heterogeneous graphs and mining graphs with relations.
    
\textbf{(v) HERec~\cite{shi2018heterogeneous}} constructs different meta-path to learn node embeddings for recommendations.

\subsection{Experiment Settings}

The performance of (micro-)video recommendation is measured by the average scores of Precision@K~\cite{fayyaz2020recommendation} that demonstrates recommendation accuracy, Normalized Discounted Cumulative Gain (NDCG@K)~\cite{articleNDCG} that describes the ranking ability, and timeliness of correctly recommended micro-videos (C-Timeliness@K)~\cite{ZHANG2017270} that measures the freshness of the recommended micro-videos, for all the users in the testing datasets.

A higher value of timeliness means that this micro-video is interacted by most of the recently viewed users. Hence, this micro-video is hot and latest. We only count the timeliness of correctly recommended micro-videos because people do not care the timeliness of the micro-videos they are not interested in.
C-Timeliness@K is expressed by Eq.~\eqref{eq:timeliness1} and Eq.~\eqref{eq:timeliness2}:
\begin{equation}
     T^{v} = \frac{1}{\left | \varphi_{G}(v,User) \right |}\sum_{u\in \varphi_{G}(v,User) }\left ( t^{uv}-t^{0} \right ),
     \label{eq:timeliness1}
\end{equation}
\begin{equation}
      T^{CR} = \frac{1}{\left | V^{CR} \right |}\sum_{v\in V^{CR}}^{} T^{v},
     \label{eq:timeliness2}
\end{equation}
where $T^{v}$ denotes the timeliness of micro-video $v$, $t^{uv}$ is the timestamp when user $u$ interacts with micro-video $v$, $t^{0}$ is a fixed timestamp selected based on datasets, and $T^{CR}$  is the timeliness of the set of correctly recommended micro-videos. 
In our experiments, $t^{0}=60,000,000$ for TikTok(1/50) and TikTok(1/5) datasets, and $t^{0}=12,000,000$ for MovieLen.

For TikTok(1/50), TikTok(1/5), and MovieLen, we randomly sample the user instances based on training rates. 
For instance, when the training rate equals 80\%, we sample 80\% users as the training set while 20\% as the testing set. The training rates are set to be 80\%, 50\%, and 20\% to simulate the situations of micro-video recommendation with different volumes of available data. 
Any user in the testing should not be the target user in the training dataset, but may join the training as a sampled neighbor of a micro-video (locate at 2 to $h$ layers in MHG). 
For testing, we sample 100 labeled latent micro-videos for each testing user, who has over 100 records in TikTok(1/50), TikTok(1/5), and MovieLen. Otherwise, we sample the maximum number of the user's records.

To better train our proposed MTHGNN model, the model parameters are initialized through Xavier distribution~\cite{glorot2010understanding} and optimized by mini-batch Adaptive Moment Estimation (Adam)~\cite{adam}. 
The various values of hyper-parameters in MTHGNN are set to fully study the model performance, including
the learning rate changed in $\left \{0.01, 0.005, 0.001, 0.0005, 0.0001 \right \}$, the batch size changed in $\left \{ 32, 64, 128 \right \}$, $\beta$ changed in $\left \{ 0.0, 0.2, 0.4, 0.6, 0.8, 1.0\right \}$, and $\eta$ changed in $\left \{ 0.05, 0.01, 0.005, 0.001 \right \}$, in which we obtain the same observations on the experiment results.
In the following, we detail and analyze the experiment results with the hyper-parameter setting listed in Table~\ref{table:parameters}.
\begin{table}[!t]
\caption{Hyper-parameter settings.}
\begin{center}
\scalebox{0.85}{
\begin{tabular}{c c c c c}
    \toprule
    \textbf{Dataset} &\textbf{Learning rate} & \textbf{Batch size} & \textbf{$\beta$} & \textbf{$\eta$} \\
    \midrule
      TikTok(1/5)  & 0.001 & 64  & 0.8 & 0.005 \\
      TikTok(1/50) & 0.001 & 64  & 0.8 & 0.005 \\ 
      MovieLen     & 0.005 & 64  & 0.2 & 0.01  \\
    \bottomrule
\end{tabular}}
\end{center}
\label{table:parameters}
\end{table}

\subsection{Comparison between MTHGNN and Baselines} \label{Performance_Comparison}
\begin{table*}
\caption{Performance comparison between MTHGNN and the baselines with training rate=80\%.}
\begin{center}
\scalebox{0.80}{
\begin{tabular}{cccccccccc}
\toprule
\multirow{2}{*}{Dataset}      & \multirow{2}{*}{Metrics} & (a)       & (b)      & (c)     & (d)              & (e)     & (f)                                  & \multicolumn{2}{l}{Improvement vs. (\%)} \\
                              &                          & GraphRec  & SASRec   & MMGCN   & R-GCN            & HERec   & MTHGNN                               & (f)-(d)             & (f)-(e)       \\
\toprule
\multirow{3}{*}{TikTok(1/50)} & Precision@10             & 0.15993   & 0.38292  & 0.50243 & 0.49094          & 0.57177 & \textbf{0.66062}                     & 34.56               & 15.54         \\
                              & NDCG@10                  & 0.47279   & 0.63534  & 0.74244 & 0.73375          & 0.78819 & \textbf{0.87735}                     & 19.57               & 11.31         \\
                              & C-Timeliness@10          & -12559836 & 601973   & 6085113 & 6342851          & 7212816 & \textbf{7448199}                     & 17.43               & 3.26          \\
\toprule
\multirow{3}{*}{TikTok(1/5)}  & Precision@10             & 0.16759   & 0.38703  & 0.50075 & 0.51505          & -       & \textbf{0.68678}                     & 33.34               & -             \\
                              & NDCG@10                  & 0.34147   & 0.62652  & 0.73588 & 0.76226          & -       & \textbf{0.90234}                     & 18.38               & -             \\
                              & C-Timeliness@10          & -15878244 & 1423568  & 6153960 & 6199042          & -       & \textbf{7295002}                     & 17.68               & -             \\
\toprule
\multirow{3}{*}{MovieLen}     & Precision@10             & 0.15488   & 0.12624  & 0.26539 & 0.33286          & 0.32806 & \textbf{0.36978}                     & 11.09               & 12.72         \\
                              & NDCG@10                  & 0.36379   & 0.26932  & 0.53340 & 0.61557          & 0.60726 & \textbf{0.64632}                     & 5.00                & 6.43          \\
                              & C-Timeliness@10          & -2636075  & -3536347 & 1021148 & \textbf{1691119} & 1303092 & 1601455                              & -5.30               & 22.90         \\        
\bottomrule
\end{tabular}
}
\end{center}
\label{table:comparison_0.8}
\end{table*} 
\begin{table*}
\caption{Performance comparison between MTHGNN and the baselines with training rate=50\%.}
\begin{center}
\scalebox{0.80}{
\begin{tabular}{cccccccccc}
\toprule
\multirow{2}{*}{Dataset}      & \multirow{2}{*}{Metrics} & (a)       & (b)      & (c)     & (d)              & (e)             & (f)              & \multicolumn{2}{l}{Improvement vs. (\%)} \\
                              &                          & GraphRec  & SASRec   & MMGCN   & R-GCN            & HERec           & MTHGNN           & (f)-(d)             & (f)-(e)            \\
\toprule
\multirow{3}{*}{TikTok(1/50)} & Precision@10             & 0.18354   & 0.37559  & 0.49051 & 0.46025          & 0.55829         & \textbf{0.62900} & 36.66               & 12.67              \\
                              & NDCG@10                  & 0.37072   & 0.62910  & 0.74061 & 0.71451          & 0.78615         & \textbf{0.85848} & 20.15               & 9.20               \\
                              & C-Timeliness@10          & -11562823 & 2092489  & 6538073 & 4921790          & 6933553         & \textbf{7475798} & 51.89               & 7.82               \\
\toprule
\multirow{3}{*}{TikTok(1/5)}  & Precision@10             & 0.16420   & 0.40178  & 0.50048 & 0.51177          & -               & \textbf{0.67617} & 32.12               & -                  \\
                              & NDCG@10                  & 0.34073   & 0.64295  & 0.73744 & 0.75775          & -               & \textbf{0.89660} & 18.32               & -                  \\
                              & C-Timeliness@10          & -17149326 & 2154885  & 6130571 & 6260566          & -               & \textbf{7358889} & 17.54               & -                  \\
\toprule
\multirow{3}{*}{MovieLen}     & Precision@10             & 0.13783   & 0.12864  & 0.26365 & 0.28965          & 0.32761         & \textbf{0.34387} & 18.72               & 4.96               \\
                              & NDCG@10                  & 0.33747   & 0.28111  & 0.53723 & 0.54147          & 0.59671         & \textbf{0.62302} & 15.06               & 4.41               \\
                              & C-Timeliness@10          & -3482251  & -3225662 & 989532  & 1139657          & \textbf{1219983} & 1185570          & -11.50              & -2.82             \\

\bottomrule
\end{tabular}
}
\end{center}
\label{table:comparison_0.5}
\end{table*} 
\begin{table*}
\caption{Performance comparison between MTHGNN and the baselines with training rate=20\%.}
\begin{center}
\scalebox{0.80}{
\begin{tabular}{cccccccccc}
\toprule

\multirow{2}{*}{Dataset}      & \multirow{2}{*}{Metrics} & (a)       & (b)      & (c)     & (d)     & (e)              & (f)              & \multicolumn{2}{l}{Improvement vs. (\%)} \\
                              &                          & GraphRec  & SASRec   & MMGCN   & R-GCN   & HERec            & MTHGNN           & (f)-(d)             & (f)-(e)            \\
\toprule
\multirow{3}{*}{TikTok(1/50)} & Precision@10             & 0.15091   & 0.37859  & 0.49023 & 0.47462 & 0.54106          & \textbf{0.58343} & 22.93               & 7.83               \\
                              & NDCG@10                  & 0.30683   & 0.62830  & 0.73994 & 0.71740 & 0.77062          & \textbf{0.82633} & 15.18               & 7.23               \\
                              & C-Timeliness@10          & -16783395 & 1639669  & 6658986 & 6019280 & 6754593          & \textbf{7387586} & 22.73               & 9.37               \\
\toprule
\multirow{3}{*}{TikTok(1/5)}  & Precision@10             & 0.16615   & 0.40549  & 0.50063 & 0.51944 & -                & \textbf{0.67175} & 29.32               & -                  \\
                              & NDCG@10                  & 0.33665   & 0.64384  & 0.73888 & 0.75420 & -                & \textbf{0.89590} & 18.79               & -                  \\
                              & C-Timeliness@10          & -13985241 & 2220509  & 6030603 & 6368694 & -                & \textbf{7368026} & 15.69               & -                  \\
\toprule
\multirow{3}{*}{MovieLen}     & Precision@10             & 0.14009   & 0.12648  & 0.26216 & 0.28737 & \textbf{0.3245}  & 0.32330          & 12.50               & -0.37              \\
                              & NDCG@10                  & 0.34488   & 0.27665  & 0.53034 & 0.55157 & 0.58965          & \textbf{0.59457} & 7.80                & 0.83               \\
                              & C-Timeliness@10          & -3110917  & -3428307 & 892244  & 600679  & \textbf{1271150} & 1086596          & 80.89                    & -14.52            \\

\bottomrule     
\end{tabular}
}
\end{center}
\label{table:comparison_0.2}
\end{table*} 

The comparison results are summarized in Table~\ref{table:comparison_0.8}, Table~\ref{table:comparison_0.5}, and Table~\ref{table:comparison_0.2}, where the performance improvement is the ratio of performance difference between MTHGNN and a baseline to the baseline performance.

In Tables~\ref{table:comparison_0.8},~\ref{table:comparison_0.5}, and~\ref{table:comparison_0.2}, MTHGNN outperforms all the baselines in terms of all metrics on TikTok(1/50) and TikTok(1/5). 
For MovieLen, MTHGNN shows nearly the state-of-the-art performance on the scores of Precision@10 and NDCG@10 in Table~\ref{table:comparison_0.8} and Table~\ref{table:comparison_0.5}, but its C-Timeliness scores are a little bit lower than those of R-GCN in Table~\ref{table:comparison_0.8} and HERec in Table~\ref{table:comparison_0.5}. 
This is because MTHGNN tries to capture the micro-videos that the users may like in the future by considering both timeliness factor and user interest. 
However, for movie recommendation in MovieLen, the users may like classic movies and do not care when the classic movies are published.  
Accordingly, MTHGNN learns user interest is more important than timeliness and thus sacrifices a little timeliness to get a more accurate recommendation of micro-videos. 
Nevertheless, the scores of C-Timeliness@10 of MTHGNN in Table~\ref{table:comparison_0.8} and Table~\ref{table:comparison_0.5} are still higher than the baselines from (a) to (c), because the LSTM and sample strategy in Algorithm~\ref{alg:BFS} of MTHGNN can still work effectively.
On the other hand, the metric scores of HERec are a little higher than those of our proposed MTHGNN on Movielen. 
The reason is that MTHGNN is a complex model containing more components compared with HERec and requires more training data to learn the model parameters. 
However, even if the training rate is only 20\%, the performance of MTHGNN is still comparable with HERec. 
%

From aspect of the characteristics of micro-videos, the properties of baseline models can be summarized as follows: (i) GraphRec considers user social behaviors, and user interests; (ii) SASRec only utilizes sequential influence and item attentions;  (iii) MMGCN, which is a model for micro-video recommendation, constructs MHG containing multiple modals; (iv) R-GCN constructs heterogeneous graphs with relation edges; and (v) HERec extracts heterogeneous meta-paths, which reveals the relations between all kinds of nodes. Our MTHGNN outperforms the baselines thanks to the multi-fold novel designs: (i) multi-modal heterogeneous graph is established to learn social relations and multi-modals; (ii) sub-graphs are built to represent frequent changed interests; (iii) relation aggregator is deployed to mine multi-type interactions; (iv) sessions attention is exploited to track the latest and historical interests; and (v) timeliness is taken into account, such as time-warping when construct MHG, LSTM aggregator, the latest session as the base to fuse interests. Therefore, MTHGNN achieves better results in terms of the three performance metrics in Table~\ref{table:comparison_0.8}, Table~\ref{table:comparison_0.5}, and Table~\ref{table:comparison_0.2}

\subsection{Ablation Study}

\begin{table*}
\centering
\caption{Ablation studies on TikTok(1/50) and MovieLen}
\begin{center}
\scalebox{0.85}{
\begin{tabular}{llllllll}
\toprule
Dataset                      & Metrics        & $\neg$Historical       & $\neg$Relation   & $\neg$All attentions   & LSTM$\rightarrow$Mean  & Single attention    & MTHGNN       \\
\midrule
\multirow{3}{*}{TikTok(1/50)} & Precision@10   & 0.54878          & 0.61149               & 0.63588                & 0.64390                & 0.64599              & \textbf{0.66062}  \\
                             & NDCG@10        & 0.82342          & 0.84523                & 0.86072                & 0.87500                & \textbf{0.88115}      & 0.87735  \\
                             & C-Timeliness@10 & \textbf{7665947} & 7380829               & 6886685                & 7225930                & 7451600              & 7448199           \\
\toprule
                             & Metrics         & LSTM$\rightarrow$Mean       & $\neg$All attention & $\neg$Historical        & $\neg$Relation  & Single attention     & MTHGNN       \\
\midrule

\multirow{3}{*}{MovieLen}    & Precision@10    & 0.30976                     & 0.31043             & 0.33791                 & 0.34230          & 0.36341             & \textbf{0.36978}  \\
                             & NDCG@10         & 0.59993                     & 0.58974             & 0.61610                 & 0.62937          & 0.64502             & \textbf{0.64632}  \\
                             & C-Timeliness@10 & 1012958                     & 816698              & 1312602                 & 1336683          & \textbf{1625869}    & 1601455           \\
\bottomrule
\end{tabular}
}
\end{center}
\label{table:anlation_study}
\end{table*}

In the ablation study, the variants of MTHGNN are compared to understand the contributions of each component developed in MTHGNN, and the results of Precision@10, NDCG@10, and C-Timeliness@10 are presented in Table~\ref{table:anlation_study}.
These variants include: 
(a) {\bf $\neg$Historical}, where all the historical sessions are removed, and only the latest session is kept; 
(b) {\bf $\neg$Relation}, where the RHMP aggregator is removed; 
(c) {\bf $\neg$All attention}, where the AHMP aggregator is removed; 
(d) {\bf Single Attention}, where $Q_{\theta }^{l}$ at each layer is removed, and a global $Q_{\theta }'$ is used instead;
(e) {\bf LSTM$\rightarrow$Mean}, where the LSTM aggregator is replaced by a mean aggregator~\cite{hamilton2017inductive}; 
and (f) {\bf MTHGNN}. 
The ablation study is conducted on TikTok(1/50) and MovieLen with the training rate at 80\%. The results of Table~\ref{table:anlation_study} are analyzed in the following.

Firstly, all of the studied components have significant contributions to the recommendation performance of our MTHGNN with different importance, which provides the flexibility to properly tailor the components of MTHGNN according to the characteristics of datasets.

Secondly, $\neg$Historical sessions are more important for micro-video recommendation than for long-movie recommendation.
Usually, a user's interest in micro-videos may be influenced by currently hot news/events and thus change frequently, while such an interest in long movies is relatively stable during a long-term period.
Therefore, the impact of temporal factor on micro-video recommendation is more significant than on long-movie recommendation.

Thirdly, the importance of the RHMP and  AHMP aggregators depends on the characteristics of datasets.
The RHMP aggregator is more crucial for micro-video recommendation on TikTok(1/50), because the multi-type interactions ({\em e.g.}, ``like'' and ``finish'') between users and micro-videos should be taken into account. 
When RHMP is removed in $\neg$Relation, the recommendation performance is reduced greatly.
On the contrary, since ``like'' is the only type of interaction in MovieLen, the AHMP aggregator becomes much more important than RHMP to mine the graph information.

Last but not least, for any target user, the number of interacted micro-videos in each session affects the performance of LSTM.
For example, the number of interacted videos in each session of each target user in TikTok(1/50) is two but in MovieLen is four.  
In TikTok(1/50), MTHGNN only aggregates two micro-video embeddings, so there is no obvious difference of aggregated embeddings between LSTM and Mean aggregator, and the replacement of LSTM in TikTok(1/50) does not reduce the scores of Precision@10 and NDCG@10 too much.
However, in MovieLen, LSTM has a longer sequence to mine a user's latent preference, and the aggregated embedding from LSTM can better reveal what the user may like in each session, which outperform Mean aggregator.
Thus, after we replace LSTM by Mean aggregator in MovieLen, the scores of Precision@10 and NDCG@10 drop to the lowest values compared with the scores of the other variants. 
The above analysis shows that with a larger number of interacted micro-videos in each session for a user, LSTM becomes more important for enhancing recommendation performance.

\subsection{Top-$K$ Recommendation}
Considering different recommendation requirements in real applications, the performance of top-$K$ recommendation is carefully investigated by changing $K$ from 1 to 10, from which our critical findings are described below.

\begin{figure}
\centering
    \subfigure[]{\includegraphics[width=0.20\textwidth]{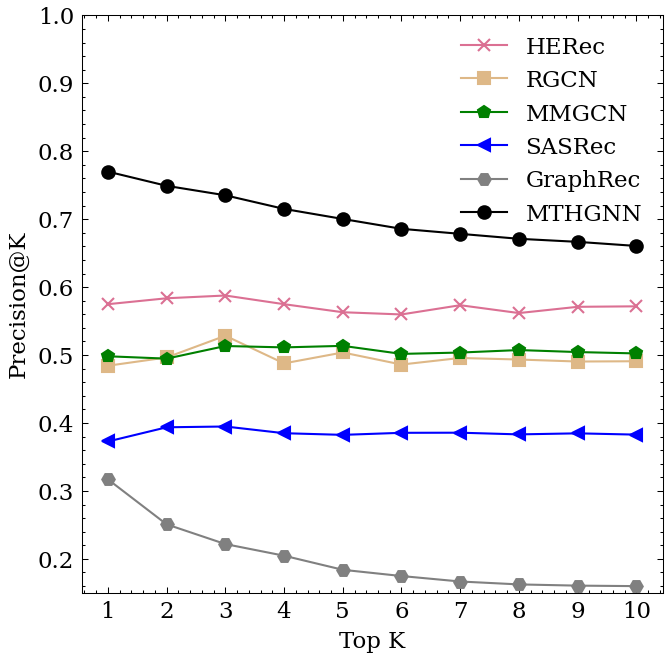}\label{fig:top1-10_precision_K} }
  \subfigure[]{\includegraphics[width=0.20\textwidth]{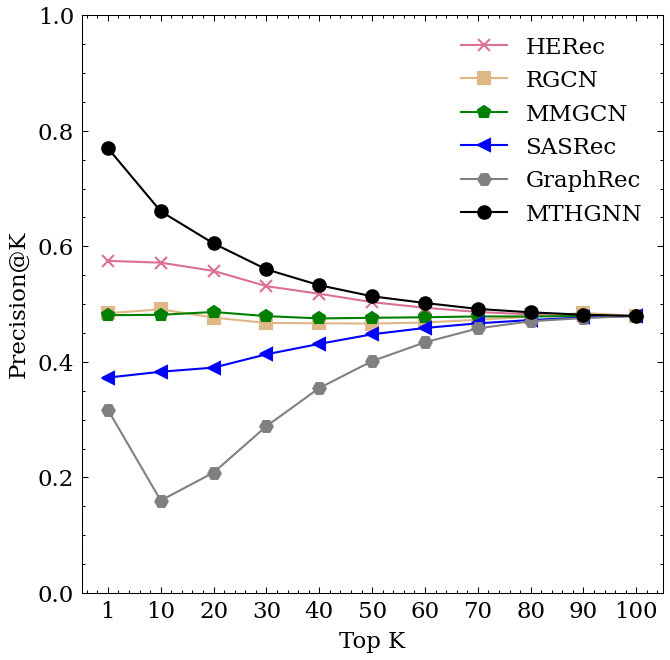}\label{fig:top1-10_precision_1_100}} 
    \subfigure[]{\includegraphics[width=0.20\textwidth]{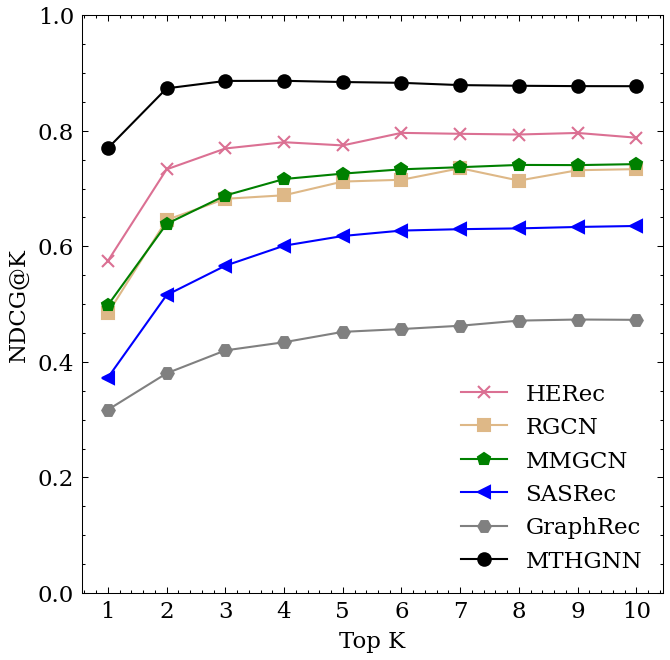}\label{fig:top1-10_ndcg}}
    \subfigure[]{\includegraphics[width=0.20\textwidth]{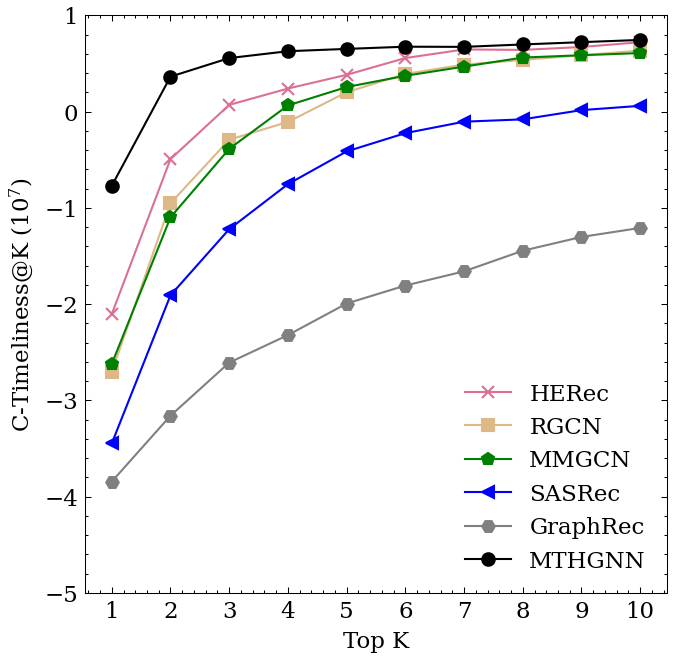}\label{fig:top1-10_ctimeliness}}
\caption{Top-$K$ recommendation performance of MTHGNN on TikTok(1/50) ((a) Precision@K, (b) Revised Precision@K, (c) NDCG@K, (d) C-Timeliness@K)}
\label{fig:top1-10}
\end{figure}

(i) The Precision@K scores on TikTok(1/50) are shown in Fig.~\ref{fig:top1-10_precision_K} and Fig.~\ref{fig:top1-10_precision_1_100}, where our MTHGNN outperforms all the baselines with $K$ increasing from 1 to 100. 
Such a performance comes from the novel development of our MTHGNN: (i) MHG is constructed to fully represent user information by using the information of nodes, relations between nodes, and heterogeneous structures; (ii) multiple aggregators are deployed to learn user preference through the relations and attentions; and (iii) the session attention is constructed to consider the user's latest and historical interest, where $\pi^{u}$ sessions are used to improve the accuracy of predicting user interest. 
For top-$K$ micro-video recommendation, in our MTHGNN mechanism, the micro-videos with higher prediction scores usually have higher probabilities of being the correct micro-videos and thus obtain higher rankings. 
As a result, when $K$ grows up, more and more micro-videos that have lower probabilities of being the correct micro-videos are recommended, leading a decrease in the Precision@K score of MTHGNN.
Especially, in Fig.~\ref{fig:top1-10_precision_1_100}, the Precision@K scores of all the compared models converge to 0.48 when $K$ is reaching to 100 that is the maximum length of a user's testing micro-video sequence, because in TikTok(1/50), the total number of correct micro-videos is limited and may be far smaller than 100.

(ii) In Fig.~\ref{fig:top1-10_ndcg}, the NDCG@K scores of MTHGNN are higher than those of the baseline models.
When $K$ is increased from 1 to 10, there is an increase in the NDCG@K scores for all the models as NDCG@K is the cumulative gain of non-negative scores of these $K$ recommended videos. 
Moreover, the increase of the NDCG@K scores follows the law of diminishing marginal benefits due to the adoption of ``$\log$'' format, which states that the micro-videos with the higher rankings receive larger weights in the cumulative gain for these $K$ recommended videos.

(iii) The results of Fig.~\ref{fig:top1-10_ctimeliness} validate that our MTHGNN can recommend the micro-videos with a higher timeliness compared with the baselines as $K$ is increased from 1 to 10.
In MTHGNN, session attention uses the embeddings of the latest micro-videos to fuse historical information based on the calculated similarities, and time-wrapping and LSTM can also help capture micro-videos with a high timeliness.
Thus, the micro-videos with a high timeliness satisfy users' latest interests and are recommended for users. 


\section{Conclusion}  \label{Conclusion}
Micro-vides own the characteristics of social relations, multi-type interactions, multi-modals, temporal factors, and timeliness.
This paper proposes a novel Multi-aggregator Time-warping Heterogeneous Graph Neural Network (MTHGNN) to address these characteristics by designing tightly coupled components, including multi-modal heterogeneous graph to simulate social relations and multi-type interactions, multi-aggregator to mine graph information from relation and attention aspects, session attention to track user interests. In this way, our MTHGNN can generate high-quality user and micro-video embeddings for accurate recommendation. 
The experiments can confirm the advantages of our MTHGNN compared with the state-of-the-arts on real datasets.

%

\bibliographystyle{ACM-Reference-Format}
\balance
\bibliography{MTHGNN_IEEE}   


\begin{thebibliography}{36}


\ifx \showCODEN    \undefined \def \showCODEN     #1{\unskip}     \fi
\ifx \showDOI      \undefined \def \showDOI       #1{#1}\fi
\ifx \showISBNx    \undefined \def \showISBNx     #1{\unskip}     \fi
\ifx \showISBNxiii \undefined \def \showISBNxiii  #1{\unskip}     \fi
\ifx \showISSN     \undefined \def \showISSN      #1{\unskip}     \fi
\ifx \showLCCN     \undefined \def \showLCCN      #1{\unskip}     \fi
\ifx \shownote     \undefined \def \shownote      #1{#1}          \fi
\ifx \showarticletitle \undefined \def \showarticletitle #1{#1}   \fi
\ifx \showURL      \undefined \def \showURL       {\relax}        \fi
\providecommand\bibfield[2]{#2}
\providecommand\bibinfo[2]{#2}
\providecommand\natexlab[1]{#1}
\providecommand\showeprint[2][]{arXiv:#2}

\bibitem[\protect\citeauthoryear{{Adomavicius} and {Tuzhilin}}{{Adomavicius}
  and {Tuzhilin}}{2005}]%
        {adomavicius2005toward}
\bibfield{author}{\bibinfo{person}{G. {Adomavicius}} {and} \bibinfo{person}{A.
  {Tuzhilin}}.} \bibinfo{year}{2005}\natexlab{}.
\newblock \showarticletitle{Toward the next generation of recommender systems:
  a survey of the state-of-the-art and possible extensions}.
\newblock \bibinfo{journal}{\emph{IEEE Transactions on Knowledge and Data
  Engineering}} \bibinfo{volume}{17}, \bibinfo{number}{6}
  (\bibinfo{year}{2005}), \bibinfo{pages}{734--749}.
\newblock


\bibitem[\protect\citeauthoryear{Busbridge, Sherburn, Cavallo, and
  Hammerla}{Busbridge et~al\mbox{.}}{2019}]%
        {busbridge2019relational}
\bibfield{author}{\bibinfo{person}{Dan Busbridge}, \bibinfo{person}{Dane
  Sherburn}, \bibinfo{person}{Pietro Cavallo}, {and} \bibinfo{person}{Nils~Y
  Hammerla}.} \bibinfo{year}{2019}\natexlab{}.
\newblock \showarticletitle{Relational graph attention networks}.
\newblock \bibinfo{journal}{\emph{arXiv preprint arXiv:1904.05811}}
  (\bibinfo{year}{2019}).
\newblock


\bibitem[\protect\citeauthoryear{Cai, Qian, Fang, and Xu}{Cai
  et~al\mbox{.}}{2021}]%
        {cai2021heterogeneous}
\bibfield{author}{\bibinfo{person}{Desheng Cai}, \bibinfo{person}{Shengsheng
  Qian}, \bibinfo{person}{Quan Fang}, {and} \bibinfo{person}{Changsheng Xu}.}
  \bibinfo{year}{2021}\natexlab{}.
\newblock \showarticletitle{Heterogeneous Hierarchical Feature Aggregation
  Network for Personalized Micro-video Recommendation}.
\newblock \bibinfo{journal}{\emph{IEEE Transactions on Multimedia}}
  (\bibinfo{year}{2021}).
\newblock


\bibitem[\protect\citeauthoryear{Corso, Cavalleri, Beaini, Li{\`o}, and
  Veli{\v{c}}kovi{\'c}}{Corso et~al\mbox{.}}{2020}]%
        {multi-aggregator}
\bibfield{author}{\bibinfo{person}{Gabriele Corso}, \bibinfo{person}{Luca
  Cavalleri}, \bibinfo{person}{Dominique Beaini}, \bibinfo{person}{Pietro
  Li{\`o}}, {and} \bibinfo{person}{Petar Veli{\v{c}}kovi{\'c}}.}
  \bibinfo{year}{2020}\natexlab{}.
\newblock \showarticletitle{Principal neighbourhood aggregation for graph
  nets}.
\newblock \bibinfo{journal}{\emph{34th Conference on Neural Information
  Processing Systems}} (\bibinfo{year}{2020}).
\newblock


\bibitem[\protect\citeauthoryear{Dataset}{Dataset}{2003}]%
        {MovieLendataset}
\bibfield{author}{\bibinfo{person}{MovieLen Dataset}.}
  \bibinfo{year}{2003}\natexlab{}.
\newblock
\newblock
\newblock
\shownote{\url {https://grouplens.org/datasets/movielens/1m}.}


\bibitem[\protect\citeauthoryear{Dataset}{Dataset}{2019}]%
        {TikTokdataset}
\bibfield{author}{\bibinfo{person}{TikTok Dataset}.}
  \bibinfo{year}{2019}\natexlab{}.
\newblock
\newblock
\newblock
\shownote{\url { https://www.biendata.xyz/competition/icmechallenge2019}.}


\bibitem[\protect\citeauthoryear{Dean}{Dean}{2021}]%
        {tiktokuseramount}
\bibfield{author}{\bibinfo{person}{Brian Dean}.}
  \bibinfo{year}{2021}\natexlab{}.
\newblock \bibinfo{title}{TikTok User Statistics (2021)}.
\newblock
\newblock
\newblock
\shownote{\url {https://backlinko.com/tiktok-users}.}


\bibitem[\protect\citeauthoryear{Fan, Ma, Li, He, Zhao, Tang, and Yin}{Fan
  et~al\mbox{.}}{2019}]%
        {fan2019graph}
\bibfield{author}{\bibinfo{person}{Wenqi Fan}, \bibinfo{person}{Yao Ma},
  \bibinfo{person}{Qing Li}, \bibinfo{person}{Yuan He}, \bibinfo{person}{Eric
  Zhao}, \bibinfo{person}{Jiliang Tang}, {and} \bibinfo{person}{Dawei Yin}.}
  \bibinfo{year}{2019}\natexlab{}.
\newblock \showarticletitle{Graph neural networks for social recommendation}.
  In \bibinfo{booktitle}{\emph{The World Wide Web Conference}}.
  \bibinfo{pages}{417--426}.
\newblock


\bibitem[\protect\citeauthoryear{Fayyaz, Ebrahimian, Nawara, Ibrahim, and
  Kashef}{Fayyaz et~al\mbox{.}}{2020}]%
        {fayyaz2020recommendation}
\bibfield{author}{\bibinfo{person}{Zeshan Fayyaz}, \bibinfo{person}{Mahsa
  Ebrahimian}, \bibinfo{person}{Dina Nawara}, \bibinfo{person}{Ahmed Ibrahim},
  {and} \bibinfo{person}{Rasha Kashef}.} \bibinfo{year}{2020}\natexlab{}.
\newblock \showarticletitle{Recommendation systems: Algorithms, challenges,
  metrics, and business opportunities}.
\newblock \bibinfo{journal}{\emph{applied sciences}} \bibinfo{volume}{10},
  \bibinfo{number}{21} (\bibinfo{year}{2020}), \bibinfo{pages}{7748}.
\newblock


\bibitem[\protect\citeauthoryear{Glorot and Bengio}{Glorot and Bengio}{2010}]%
        {glorot2010understanding}
\bibfield{author}{\bibinfo{person}{Xavier Glorot} {and} \bibinfo{person}{Yoshua
  Bengio}.} \bibinfo{year}{2010}\natexlab{}.
\newblock \showarticletitle{Understanding the difficulty of training deep
  feedforward neural networks}. In \bibinfo{booktitle}{\emph{Proceedings of the
  thirteenth international conference on artificial intelligence and
  statistics}}. JMLR Workshop and Conference Proceedings,
  \bibinfo{pages}{249--256}.
\newblock


\bibitem[\protect\citeauthoryear{Hamilton, Ying, and Leskovec}{Hamilton
  et~al\mbox{.}}{2017}]%
        {hamilton2017inductive}
\bibfield{author}{\bibinfo{person}{William~L Hamilton}, \bibinfo{person}{Rex
  Ying}, {and} \bibinfo{person}{Jure Leskovec}.}
  \bibinfo{year}{2017}\natexlab{}.
\newblock \showarticletitle{Inductive representation learning on large graphs}.
  In \bibinfo{booktitle}{\emph{Proceedings of the 31st International Conference
  on Neural Information Processing Systems}}. \bibinfo{pages}{1025--1035}.
\newblock


\bibitem[\protect\citeauthoryear{Hsu and Li}{Hsu and Li}{2021}]%
        {hsu2021retagnn}
\bibfield{author}{\bibinfo{person}{Cheng Hsu} {and} \bibinfo{person}{Cheng-Te
  Li}.} \bibinfo{year}{2021}\natexlab{}.
\newblock \showarticletitle{RetaGNN: Relational Temporal Attentive Graph Neural
  Networks for Holistic Sequential Recommendation}. In
  \bibinfo{booktitle}{\emph{Proceedings of the Web Conference 2021}}.
  \bibinfo{pages}{2968--2979}.
\newblock


\bibitem[\protect\citeauthoryear{Jiang, Wang, Wei, Gao, Wang, and Nie}{Jiang
  et~al\mbox{.}}{2020}]%
        {jiang2020aspect}
\bibfield{author}{\bibinfo{person}{Hao Jiang}, \bibinfo{person}{Wenjie Wang},
  \bibinfo{person}{Yinwei Wei}, \bibinfo{person}{Zan Gao},
  \bibinfo{person}{Yinglong Wang}, {and} \bibinfo{person}{Liqiang Nie}.}
  \bibinfo{year}{2020}\natexlab{}.
\newblock \showarticletitle{What Aspect Do You Like: Multi-scale Time-aware
  User Interest Modeling for Micro-video Recommendation}. In
  \bibinfo{booktitle}{\emph{Proceedings of the 28th ACM International
  Conference on Multimedia}}. \bibinfo{pages}{3487--3495}.
\newblock


\bibitem[\protect\citeauthoryear{Kang and McAuley}{Kang and McAuley}{2018}]%
        {kang2018self}
\bibfield{author}{\bibinfo{person}{Wang-Cheng Kang} {and}
  \bibinfo{person}{Julian McAuley}.} \bibinfo{year}{2018}\natexlab{}.
\newblock \showarticletitle{Self-attentive sequential recommendation}. In
  \bibinfo{booktitle}{\emph{2018 IEEE International Conference on Data Mining
  (ICDM)}}. IEEE, \bibinfo{pages}{197--206}.
\newblock


\bibitem[\protect\citeauthoryear{Kingma and Ba}{Kingma and Ba}{2014}]%
        {adam}
\bibfield{author}{\bibinfo{person}{Diederik Kingma} {and}
  \bibinfo{person}{Jimmy Ba}.} \bibinfo{year}{2014}\natexlab{}.
\newblock \showarticletitle{Adam: A Method for Stochastic Optimization}.
\newblock \bibinfo{journal}{\emph{International Conference on Learning
  Representations}} (\bibinfo{date}{12} \bibinfo{year}{2014}).
\newblock


\bibitem[\protect\citeauthoryear{Kwai}{Kwai}{2022}]%
        {Kwaiwebsite}
\bibfield{author}{\bibinfo{person}{Kwai}.} \bibinfo{year}{2022}\natexlab{}.
\newblock
\newblock
\newblock
\shownote{\url { https://www.kwai.com/}.}


\bibitem[\protect\citeauthoryear{Liu and Chen}{Liu and Chen}{2019}]%
        {8784862}
\bibfield{author}{\bibinfo{person}{Shang Liu} {and} \bibinfo{person}{Zhenzhong
  Chen}.} \bibinfo{year}{2019}\natexlab{}.
\newblock \showarticletitle{Sequential Behavior Modeling for Next Micro-Video
  Recommendation with Collaborative Transformer}. In
  \bibinfo{booktitle}{\emph{2019 IEEE International Conference on Multimedia
  and Expo (ICME)}}. \bibinfo{pages}{460--465}.
\newblock


\bibitem[\protect\citeauthoryear{Liu, Liu, Tian, Wang, Niu, Song, and Li}{Liu
  et~al\mbox{.}}{2021}]%
        {liu2021concept}
\bibfield{author}{\bibinfo{person}{Yiyu Liu}, \bibinfo{person}{Qian Liu},
  \bibinfo{person}{Yu Tian}, \bibinfo{person}{Changping Wang},
  \bibinfo{person}{Yanan Niu}, \bibinfo{person}{Yang Song}, {and}
  \bibinfo{person}{Chenliang Li}.} \bibinfo{year}{2021}\natexlab{}.
\newblock \showarticletitle{Concept-Aware Denoising Graph Neural Network for
  Micro-Video Recommendation}. In \bibinfo{booktitle}{\emph{Proceedings of the
  30th ACM International Conference on Information \& Knowledge Management}}.
  \bibinfo{pages}{1099--1108}.
\newblock


\bibitem[\protect\citeauthoryear{Lu, Huang, Zhang, Han, Chen, Zhao, and Wu}{Lu
  et~al\mbox{.}}{2021}]%
        {lu2021multi}
\bibfield{author}{\bibinfo{person}{Yujie Lu}, \bibinfo{person}{Yingxuan Huang},
  \bibinfo{person}{Shengyu Zhang}, \bibinfo{person}{Wei Han},
  \bibinfo{person}{Hui Chen}, \bibinfo{person}{Zhou Zhao}, {and}
  \bibinfo{person}{Fei Wu}.} \bibinfo{year}{2021}\natexlab{}.
\newblock \showarticletitle{Multi-trends Enhanced Dynamic Micro-video
  Recommendation}.
\newblock \bibinfo{journal}{\emph{arXiv preprint arXiv:2110.03902}}
  (\bibinfo{year}{2021}).
\newblock


\bibitem[\protect\citeauthoryear{Ma, Li, Zhong, Zhao, Zhu, and Li}{Ma
  et~al\mbox{.}}{2018}]%
        {ma2018lga}
\bibfield{author}{\bibinfo{person}{Jingwei Ma}, \bibinfo{person}{Guang Li},
  \bibinfo{person}{Mingyang Zhong}, \bibinfo{person}{Xin Zhao},
  \bibinfo{person}{Lei Zhu}, {and} \bibinfo{person}{Xue Li}.}
  \bibinfo{year}{2018}\natexlab{}.
\newblock \showarticletitle{LGA: latent genre aware micro-video recommendation
  on social media}.
\newblock \bibinfo{journal}{\emph{Multimedia Tools and Applications}}
  \bibinfo{volume}{77}, \bibinfo{number}{3} (\bibinfo{year}{2018}),
  \bibinfo{pages}{2991--3008}.
\newblock


\bibitem[\protect\citeauthoryear{Ma, Wen, Zhong, Chen, and Li}{Ma
  et~al\mbox{.}}{2019}]%
        {ma2019mmm}
\bibfield{author}{\bibinfo{person}{Jingwei Ma}, \bibinfo{person}{Jiahui Wen},
  \bibinfo{person}{Mingyang Zhong}, \bibinfo{person}{Weitong Chen}, {and}
  \bibinfo{person}{Xue Li}.} \bibinfo{year}{2019}\natexlab{}.
\newblock \showarticletitle{MMM: multi-source multi-net micro-video
  recommendation with clustered hidden item representation learning}.
\newblock \bibinfo{journal}{\emph{Data Science and Engineering}}
  \bibinfo{volume}{4}, \bibinfo{number}{3} (\bibinfo{year}{2019}),
  \bibinfo{pages}{240--253}.
\newblock


\bibitem[\protect\citeauthoryear{Maas, Hannun, Ng, et~al\mbox{.}}{Maas
  et~al\mbox{.}}{2013}]%
        {maas2013rectifier}
\bibfield{author}{\bibinfo{person}{Andrew~L Maas}, \bibinfo{person}{Awni~Y
  Hannun}, \bibinfo{person}{Andrew~Y Ng}, {et~al\mbox{.}}}
  \bibinfo{year}{2013}\natexlab{}.
\newblock \showarticletitle{Rectifier nonlinearities improve neural network
  acoustic models}. In \bibinfo{booktitle}{\emph{Proc. ICML}}. Citeseer,
  \bibinfo{pages}{3}.
\newblock


\bibitem[\protect\citeauthoryear{Nair and Hinton}{Nair and Hinton}{2010}]%
        {nair2010rectified}
\bibfield{author}{\bibinfo{person}{Vinod Nair} {and}
  \bibinfo{person}{Geoffrey~E Hinton}.} \bibinfo{year}{2010}\natexlab{}.
\newblock \showarticletitle{Rectified linear units improve restricted boltzmann
  machines}. In \bibinfo{booktitle}{\emph{Proceedings of the 27th International
  Conference on International Conference on Machine Learning}}.
\newblock


\bibitem[\protect\citeauthoryear{Rendle, Freudenthaler, Gantner, and
  Schmidt-Thieme}{Rendle et~al\mbox{.}}{2009}]%
        {BPR}
\bibfield{author}{\bibinfo{person}{Steffen Rendle}, \bibinfo{person}{Christoph
  Freudenthaler}, \bibinfo{person}{Zeno Gantner}, {and} \bibinfo{person}{Lars
  Schmidt-Thieme}.} \bibinfo{year}{2009}\natexlab{}.
\newblock \showarticletitle{BPR: Bayesian Personalized Ranking from Implicit
  Feedback}. In \bibinfo{booktitle}{\emph{Proceedings of the Twenty-Fifth
  Conference on Uncertainty in Artificial Intelligence}}.
  \bibinfo{pages}{452–461}.
\newblock


\bibitem[\protect\citeauthoryear{Schlichtkrull, Kipf, Bloem, Van Den~Berg,
  Titov, and Welling}{Schlichtkrull et~al\mbox{.}}{2018}]%
        {rgcn}
\bibfield{author}{\bibinfo{person}{Michael Schlichtkrull},
  \bibinfo{person}{Thomas~N Kipf}, \bibinfo{person}{Peter Bloem},
  \bibinfo{person}{Rianne Van Den~Berg}, \bibinfo{person}{Ivan Titov}, {and}
  \bibinfo{person}{Max Welling}.} \bibinfo{year}{2018}\natexlab{}.
\newblock \showarticletitle{Modeling relational data with graph convolutional
  networks}. In \bibinfo{booktitle}{\emph{European semantic web conference}}.
  Springer, \bibinfo{pages}{593--607}.
\newblock


\bibitem[\protect\citeauthoryear{Shi, Hu, Zhao, and Philip}{Shi
  et~al\mbox{.}}{2018}]%
        {shi2018heterogeneous}
\bibfield{author}{\bibinfo{person}{Chuan Shi}, \bibinfo{person}{Binbin Hu},
  \bibinfo{person}{Wayne~Xin Zhao}, {and} \bibinfo{person}{S~Yu Philip}.}
  \bibinfo{year}{2018}\natexlab{}.
\newblock \showarticletitle{Heterogeneous information network embedding for
  recommendation}.
\newblock \bibinfo{journal}{\emph{IEEE Transactions on Knowledge and Data
  Engineering}} \bibinfo{volume}{31}, \bibinfo{number}{2}
  (\bibinfo{year}{2018}), \bibinfo{pages}{357--370}.
\newblock


\bibitem[\protect\citeauthoryear{Tian, Liu, Sun, Jiang, and Zhu}{Tian
  et~al\mbox{.}}{2021}]%
        {groupgraph}
\bibfield{author}{\bibinfo{person}{Zhiqiang Tian}, \bibinfo{person}{Yezheng
  Liu}, \bibinfo{person}{Jianshan Sun}, \bibinfo{person}{Yuanchun Jiang}, {and}
  \bibinfo{person}{Mingyue Zhu}.} \bibinfo{year}{2021}\natexlab{}.
\newblock \showarticletitle{Exploiting Group Information for Personalized
  Recommendation with Graph Neural Networks}.
\newblock \bibinfo{journal}{\emph{ACM Transactions on Information Systems}}
  \bibinfo{volume}{40}, \bibinfo{number}{2}, Article \bibinfo{articleno}{27}
  (\bibinfo{year}{2021}), \bibinfo{numpages}{23}~pages.
\newblock


\bibitem[\protect\citeauthoryear{TikTok}{TikTok}{2022}]%
        {tiktokwebsite}
\bibfield{author}{\bibinfo{person}{TikTok}.} \bibinfo{year}{2022}\natexlab{}.
\newblock
\newblock
\newblock
\shownote{\url { https://www.tiktok.com/}.}


\bibitem[\protect\citeauthoryear{Veli{\v{c}}kovi{\'c}, Cucurull, Casanova,
  Romero, Lio, and Bengio}{Veli{\v{c}}kovi{\'c} et~al\mbox{.}}{2017}]%
        {velivckovic2017graph}
\bibfield{author}{\bibinfo{person}{Petar Veli{\v{c}}kovi{\'c}},
  \bibinfo{person}{Guillem Cucurull}, \bibinfo{person}{Arantxa Casanova},
  \bibinfo{person}{Adriana Romero}, \bibinfo{person}{Pietro Lio}, {and}
  \bibinfo{person}{Yoshua Bengio}.} \bibinfo{year}{2017}\natexlab{}.
\newblock \showarticletitle{Graph attention networks}.
\newblock \bibinfo{journal}{\emph{arXiv preprint arXiv:1710.10903}}
  (\bibinfo{year}{2017}).
\newblock


\bibitem[\protect\citeauthoryear{Wang, Girshick, Gupta, and He}{Wang
  et~al\mbox{.}}{2018}]%
        {wang2018non}
\bibfield{author}{\bibinfo{person}{Xiaolong Wang}, \bibinfo{person}{Ross
  Girshick}, \bibinfo{person}{Abhinav Gupta}, {and} \bibinfo{person}{Kaiming
  He}.} \bibinfo{year}{2018}\natexlab{}.
\newblock \showarticletitle{Non-local neural networks}. In
  \bibinfo{booktitle}{\emph{Proceedings of the IEEE conference on computer
  vision and pattern recognition}}. \bibinfo{pages}{7794--7803}.
\newblock


\bibitem[\protect\citeauthoryear{Wang, Wang, Li, He, Liu, and Chen}{Wang
  et~al\mbox{.}}{2013}]%
        {articleNDCG}
\bibfield{author}{\bibinfo{person}{Yining Wang}, \bibinfo{person}{Liwei Wang},
  \bibinfo{person}{Yuanzhi Li}, \bibinfo{person}{Di He},
  \bibinfo{person}{Tie-Yan Liu}, {and} \bibinfo{person}{Wei Chen}.}
  \bibinfo{year}{2013}\natexlab{}.
\newblock \showarticletitle{A Theoretical Analysis of NDCG Type Ranking
  Measures}.
\newblock \bibinfo{journal}{\emph{Journal of Machine Learning Research}}
  \bibinfo{volume}{30} (\bibinfo{date}{04} \bibinfo{year}{2013}).
\newblock


\bibitem[\protect\citeauthoryear{Wei, Wang, Nie, He, Hong, and Chua}{Wei
  et~al\mbox{.}}{2019}]%
        {wei2019mmgcn}
\bibfield{author}{\bibinfo{person}{Yinwei Wei}, \bibinfo{person}{Xiang Wang},
  \bibinfo{person}{Liqiang Nie}, \bibinfo{person}{Xiangnan He},
  \bibinfo{person}{Richang Hong}, {and} \bibinfo{person}{Tat-Seng Chua}.}
  \bibinfo{year}{2019}\natexlab{}.
\newblock \showarticletitle{MMGCN: Multi-modal graph convolution network for
  personalized recommendation of micro-video}. In
  \bibinfo{booktitle}{\emph{Proceedings of the 27th ACM International
  Conference on Multimedia}}. \bibinfo{pages}{1437--1445}.
\newblock


\bibitem[\protect\citeauthoryear{Yao, Zhang, Zhao, Fan, Zhu, He, and Wu}{Yao
  et~al\mbox{.}}{2021}]%
        {AAAImultiinteraction}
\bibfield{author}{\bibinfo{person}{Dong Yao}, \bibinfo{person}{Shengyu Zhang},
  \bibinfo{person}{Zhou Zhao}, \bibinfo{person}{Wenyan Fan},
  \bibinfo{person}{Jieming Zhu}, \bibinfo{person}{Xiuqiang He}, {and}
  \bibinfo{person}{Fei Wu}.} \bibinfo{year}{2021}\natexlab{}.
\newblock \showarticletitle{Modeling High-order Interactions across
  Multi-interests for Micro-video Reommendation}.
\newblock \bibinfo{journal}{\emph{AAAI 2021}} (\bibinfo{date}{04}
  \bibinfo{year}{2021}).
\newblock


\bibitem[\protect\citeauthoryear{Zhang, Liu, and Zeng}{Zhang
  et~al\mbox{.}}{2017}]%
        {ZHANG2017270}
\bibfield{author}{\bibinfo{person}{Fuguo Zhang}, \bibinfo{person}{Qihua Liu},
  {and} \bibinfo{person}{An Zeng}.} \bibinfo{year}{2017}\natexlab{}.
\newblock \showarticletitle{Timeliness in recommender systems}.
\newblock \bibinfo{journal}{\emph{Expert Systems with Applications}}
  \bibinfo{volume}{85} (\bibinfo{year}{2017}), \bibinfo{pages}{270--278}.
\newblock


\bibitem[\protect\citeauthoryear{Zhang, Yao, Sun, and Tay}{Zhang
  et~al\mbox{.}}{2019}]%
        {surveyRS}
\bibfield{author}{\bibinfo{person}{Shuai Zhang}, \bibinfo{person}{Lina Yao},
  \bibinfo{person}{Aixin Sun}, {and} \bibinfo{person}{Yi Tay}.}
  \bibinfo{year}{2019}\natexlab{}.
\newblock \showarticletitle{Deep Learning Based Recommender System: A Survey
  and New Perspectives}.
\newblock \bibinfo{journal}{\emph{ACM Comput. Surv.}} \bibinfo{volume}{52},
  \bibinfo{number}{1}, Article \bibinfo{articleno}{5} (\bibinfo{date}{Feb}
  \bibinfo{year}{2019}), \bibinfo{numpages}{38}~pages.
\newblock


\bibitem[\protect\citeauthoryear{Zhao, Cheng, Hong, and Chi}{Zhao
  et~al\mbox{.}}{2015}]%
        {zhao2015improving}
\bibfield{author}{\bibinfo{person}{Zhe Zhao}, \bibinfo{person}{Zhiyuan Cheng},
  \bibinfo{person}{Lichan Hong}, {and} \bibinfo{person}{Ed~H Chi}.}
  \bibinfo{year}{2015}\natexlab{}.
\newblock \showarticletitle{Improving user topic interest profiles by behavior
  factorization}. In \bibinfo{booktitle}{\emph{Proceedings of the 24th
  International Conference on World Wide Web}}. \bibinfo{pages}{1406--1416}.
\newblock


\end{thebibliography}

\end{sloppypar}
\end{document}